\documentclass[prb,twocolumn,showpacs,amsmath,amssymb,superscriptaddress]{revtex4}
\usepackage{graphicx}
\usepackage{dcolumn}
\usepackage{bm}
\usepackage{amssymb}
\usepackage{dcolumn}
\usepackage{xcolor}

\begin{document}

\title{A stable path to ferromagnetic hydrogenated graphene growth}
\author{Shayan Hemmatiyan}
\affiliation{Department of Physics, Texas A\&M University, College Station, TX 77843-4242, USA}
\author{Marco Polini}
\affiliation{NEST, Istituto Nanoscienze - CNR and Scuola Normale Superiore, I-56126 Pisa, Italy}
\author{Artem Abanov}
\affiliation{Department of Physics, Texas A\&M University, College Station, TX 77843-4242, USA}
\author{Allan H. MacDonald }
\affiliation{Department of Physics, University of Texas at Austin, Austin, Texas 78712-1081, USA}
\author{Jairo Sinova}
\affiliation{Institut f\"ur Physik, Johannes Gutenberg Universit\"at Mainz, D-55099 Mainz, Germany}
\affiliation{Department of Physics, Texas A\&M University, College Station, TX 77843-4242, USA}
\affiliation{Institute of Physics ASCR, Cukrovarnick\'a 10, 162 53 Praha 6, Czech Republic}
\date{\today}
\begin{abstract}
In this paper, we propose a practical way to stabilize half-hydrogenated graphene (graphone). We show that the dipole moments induced by a hexagonal-boron nitride (h-BN) substrate on graphene stabilize the hydrogen atoms on one sublattice of the graphene layer and suppress the migration of the adsorbed hydrogen atoms. Based upon first principle spin polarized density of states (DOS) calculations, we show that the graphone obtained in different graphene/h-BN heterostructures exhibits a half metallic state. We propose to use this new exotic material for spin valve systems and other spintronics devices.
\end{abstract}
\pacs{}

\maketitle

\section{Introduction}
The low electronic density of states near the fermi energy and the atomic thickness of graphene make it a very attractive material for high frequency large-scale integrated electronics \cite{geim2007rise}. However, due to its semi-metallic nature (zero band-gap at neutrality point), graphene exhibits a small ON/OFF switching ratio ($<$10 at room temperature). This problem inhibits the application of graphene for charge based logic devices and integrated circuits. \cite{britnell2012field, hemmatiyan2013vertical, pan2013vertical} 

One possibility to overcome the small ON/OFF switch ratio is to control charge current via spin. The proposed mechanisms to adjust the current include applying an external magnetic field or a magnetization switching (charge based) via spin polarized current in spin valve systems and tunneling magneto resistance (TMR) devices.\cite{kiselev2003microwave, bertotti2005magnetization, PhysRevLett.100.186805, PhysRevLett.74.3273}. The magnetic properties of graphene have also been extensively studied
in the recent years.\cite{sorella2007semi, PhysRevB.75.125408, yazyev2010emergence, PhysRevB.77.195428, PhysRevB.78.235435, PhysRevB.77.134114, PhysRevB.84.125410} 
It has been shown, that pure graphene exhibits only weak antiferromagnetic (AF) order \cite{sorella2007semi} at near room temperature. However, studies of structural and other defects show that induced  $sp^3$-type hybridization, such as mono-vacancies create a local spin moment and, in certain circumstances, promote a robust long range magnetic order \cite{PhysRevB.75.125408,yazyev2010emergence}.

A similar effect is obtained by chemical functionalization of the graphene with elements such as hydrogen.\cite{PhysRevB.75.125408, yazyev2010emergence, zhou2009ferromagnetism, moaied2014hydrogenation} 
Maximum magnetic moment (1$\mu_B$ per cell) is predicted for the half hydrogenated-graphene (graphone). In such theoretical systems the hydrogen atoms are adsorbed by only one graphene sublattice. Induced $sp^{3}$ hybridization then results in localized magnetic moment in the same way as for defects. However, the overlap between the $p_{z}$ orbitals of the nearest carbons  is sufficient to create  a long range magnetic order at room temperature \cite{yazyev2010emergence}. The $p_z$ orbitals have a near $3$ eV band gap \cite{PhysRevB.82.153404, kharche2011quasiparticle} in graphone.

Although graphone could make a breakthrough in spintronics, its fabrication is an experimentally challenging task. Fundamentally, there are two obstacles:
i) the symmetry between sublattices, and ii) the lack of barrier for the trapped hydrogen atoms to migrate between the sublattices. \cite{boukhvalov2010stable}

Recently, there has been some progress in fabrication of partially hydrogenated graphene \cite{peng2014new, giesbers2013interface} under certain experimental conditions (i.e low temperature) \cite{peng2014new}. Zhou et al proposed a functionalized heterostructures of fully hydrogenated graphene (graphane) on the top of hexagonal-boron nitride (h-BN) \cite{zhou2012fabricate}. The idea is creating an active nitrogen agent by exposing the system to fluorine. The instability of nitrogen-fluorine bond will increase the electronegativity of nitrogen thus creating an active nitrogen site. By applying pressure on the fluorinated h-BN layer, the system undergoes a structural transition from graphane to semi-hydrogenated graphene by adsorption of all the hydrogen atoms from one sublattice in graphane to h-BN layer \cite{zhou2012fabricate}. 

Hexagonal boron nitride is the insulating isomorph of graphite (honeycomb lattice with boron and nitrogen on two adjacent sublattices) with a large band gap of 5.97 eV. It was shown \cite{dean2010boron} that it is a superior substrate for graphene for homogenous and high quality graphene fabrication \cite{dean2010boron, kim2013synthesis, yang2013epitaxial}. The small lattice mismatching ($1.7$\%) and atomically planer structure of h-BN (free of dangling bonds and charge traps) preserve properties of graphene such as charge carrier mobility.

In this paper, we present a practical method to solve the two obstacles (symmetry of two sublattices and mobility of hydrogen) using the functionalized graphene hybrid structure with h-BN. Our proposed experimental set up includes two steps: fabrication of graphene on h-BN substrate then exposing the system into the hydrogen plasma. The electrical dipole induced by the substrate in addition to small buckling of carbon bonds will trap hydrogen in one sublattice and will kinematically stabilize the system.

We show that for h-BN the difference in electronegativity of nitrogen and boron creates a dipole moment for each nitrogen site. This dipole moment breaks the equivalency of two carbon atoms in two different graphene sublattices. The similar screening effect has been reported in multilayer graphene but with different strength \cite{moaied2014theoretical}. Moreover, the screening effect of h-BN will generate a buckling in the graphene layer. This buckling will change the vertical position of the one sublattice with respect to the other sublattice and will enhance the coverage rate of the hydrogen in one sublattice. The dipole moment is also responsible for the increased migration barrier in the adsorbed hydrogen atoms, effectively pinning the hydrogen atoms to one sublattice.
 
We also show that the very same dipole moment induced by the h-BN substrate changes the fermi energy of the graphone layer and decreases the band gap of the graphone from near 3 eV in pristine graphone to 1.93 eV in graphone/h-BN heterostructure. \cite{kharche2011quasiparticle}

\begin{figure}[h]
\centering
\includegraphics[width=0.75\columnwidth]{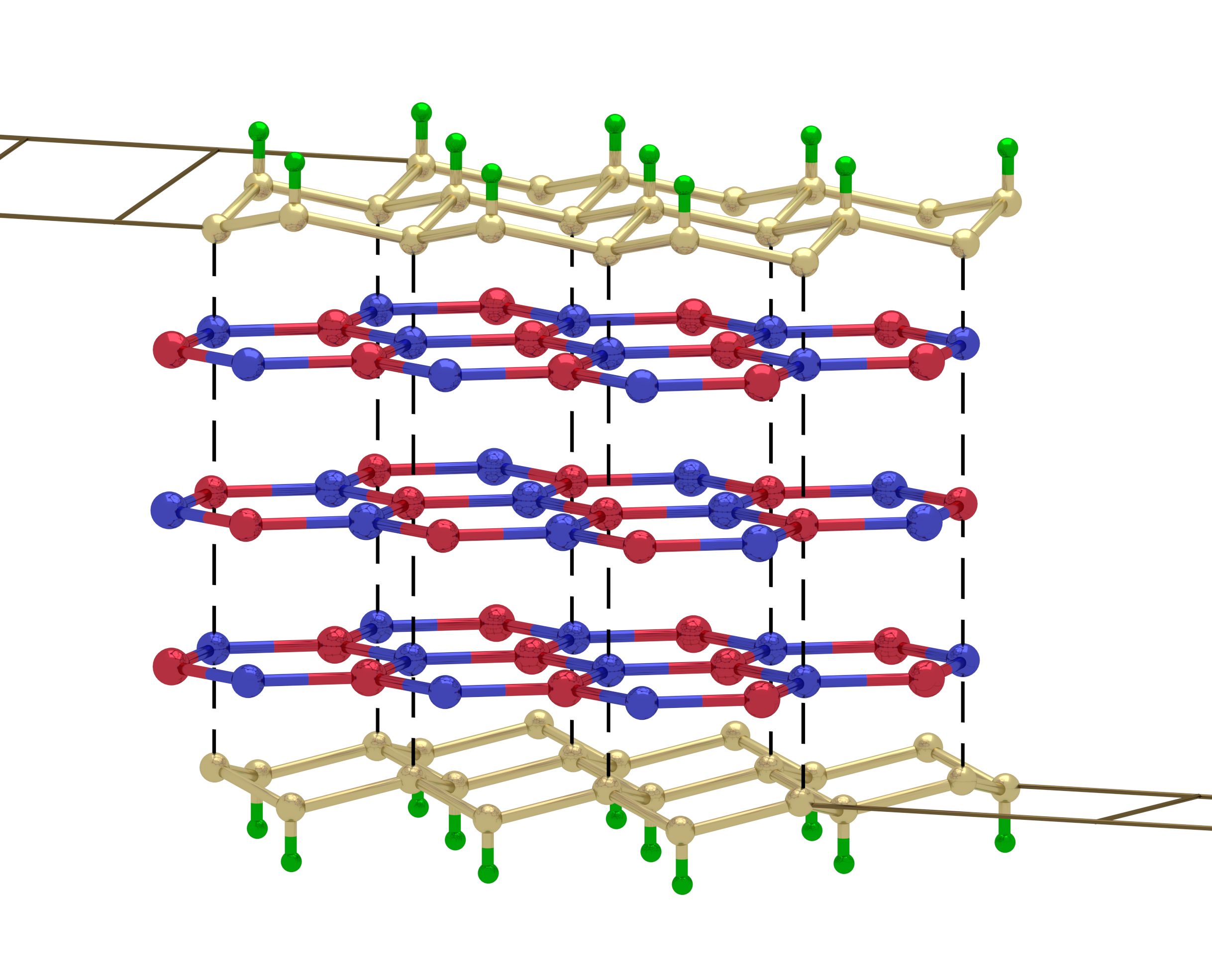}
\caption{TMR transistor: Graphone(B)/h-BN/Graphone(B), two ferromagnetic layers of graphone and few (three) layer of h-BN between} 
\label{fig:tmr3bn}
\end{figure}

We also show that the spin-orbit coupling (SOC) in the graphone increases due to a change in hybridization. In graphene, carbon has small SOC due to its $sp^2$ hybridization. The $d$ orbital has the main contribution in creating a 24 $\mu eV$ intrinsic band-gap \cite{PhysRevB.82.245412}. However, it has been seen that the hydrogen atom enhances spin orbit interactions in graphene via transition from $sp^2$ to $sp^3$ type hybridization due to bending of bonds between carbon atoms  \cite{zhou2010enhanced, PhysRevLett.110.246602}. We show this effect will increase in the presence of the h-BN.

Finally, we propose to use the graphone/h-BN heterostructures in TMR devices as shown in Fig. \ref{fig:tmr3bn}. 
The tunneling current is tuned by changing the magnetization of the one of the electrodes. \cite{PhysRevLett.74.3273, kiselev2003microwave, bertotti2005magnetization, PhysRevLett.100.186805}
We show that multilayer graphone/h-BN heterostructure is a half metal with near 100\% spin polarization. It was speculated before \cite{julliere1975tunneling} that such materials are ideal for the TMR devices.

This paper is organized as follows. In \ref{sec:methods} we describe the methods we use. In section \ref{sec:stacking} we show how the screening effect of h-BN depends on stacking. In section \ref{sec:stability}, we calculate the mobility of hydrogen atoms on both pristine graphene and graphene/h-BN heterostructure. In section \ref{sec:stability}, we show that the graphone/h-BN heterostructure is stable. In section \ref{sec:Magnetic}, we perform spin polarized calculations, determine the stable magnetic state, and the magnitude of the SOC. In section \ref{sec:TMR}, we show that not only h-BN heterostructure can be
used to trap hydrogen for fabrication of graphone, but could also be
utilized as an insulator in TMR devices.

\section{Methods}
\label{sec:methods}
All calculations have been done within the first principle framework of the {\sc Quantum ESPRESSO} package \cite{QE-2009}. We utilize both the local density  \cite{PhysRevB.23.5048} (LDA) and the generalized gradient approximations \cite{PhysRevLett.77.3865} (GGA) for graphone on top of h-BN. In addition to GGA, van der Waals (VDW) interactions have been treated through VDW-DF2 code \cite{langreth2009density} within the {\sc Quantum ESPRESSO} package. 

It was shown \cite{PhysRevB.77.035427} previously, that LDA is accurate for calculations of the interlayer binding in graphite and multi-layer graphene, while GGA  better matches the experimental data for intra-band interactions and structure. We note, however, that the discrepancy between the two methods is small and is not essential for the main results of our paper.

LDA calculations are done under ultra-soft, norm-conserving Perdew-Zunger \cite{PhysRevB.23.5048} (PZ) exchange correlation with the energy cut-off of 60 Ry. For finding optimized structure and activation energy, we used a large supercell, $5 \times 5$ unit cells, to prevent overlapping between distorted areas. Also, a 20 {\AA} vacuum space was used to avoid interactions between two periodic layers. 

To find the minimum-energy paths through the migration barrier for the hydrogen atoms, the strong distortion of covalent bonds must be taken into account.
For this purpose, we used nudged-elastic-band (NEB) method outlined in Reference \onlinecite{boukhvalov2010stable}.

For finding the optimized structure, we used conjugate-gradient (CG) method with the force and energy convergence parameters $10^{-3}$ Ry/a.u. and $10^{-5}$ Ry, respectively. A mesh of $4\times4\times1$ k-points in the Mokhorst-Park method \cite{monkhorst1976special} and a cold smearing \cite{PhysRevLett.82.3296} of 0.01 eV degauss was implemented. 

For the calculation of the magnetic properties, we used a spin-polarized LDA (LSDA) with a fully relativistic and norm-conserving Perdew-Burke-Ernzerholf (PBE) method \cite{PhysRevLett.77.3865} which has non-linear core corrections; energy and charge density cut-offs were 50 eV and 500 eV, respectively. We used a non-collinear calculation in the presence of SOC with all spins constrained along the z-direction. Our supercell consists of $2\times2$ unit-cells. 

To find the SOC, we fit the band structure of graphone/h-BN with SOC to the tight binding model near the Dirac point. A marzari-vanderbilt (m-v) smearing of 0.01 eV has been used for calculating density of states. The force and energy optimization accuracy  was $10^{-3}$ Ry/a.u. and $10^{-5}$ Ry, respectively. 

\section{Stacking graphene on h-BN}
\label{sec:stacking}
In this section, we discuss the different ways of stacking in the graphene/h-BN layers, in which both graphene and h-BN have the hexagonal lattices. The graphene has carbon atoms in both sublattices, while h-BN has boron in one sublattice and nitrogen in the other.

\begin{figure}[h]
\centering
\includegraphics[width=0.82\columnwidth]{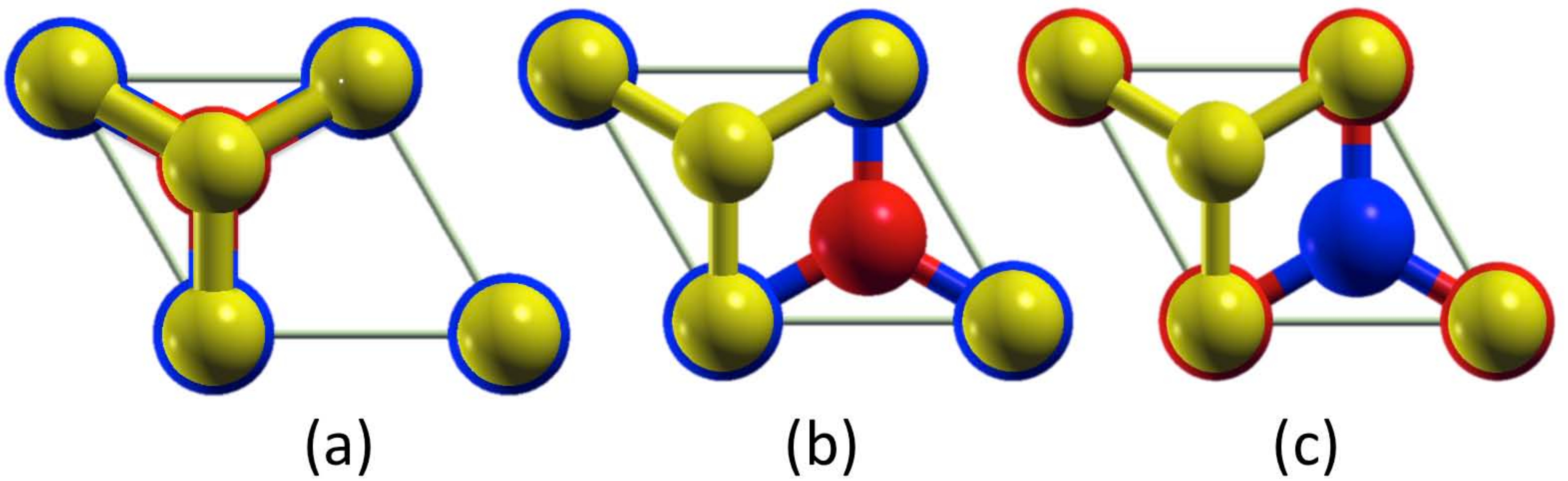}
\includegraphics[width=0.18\columnwidth]{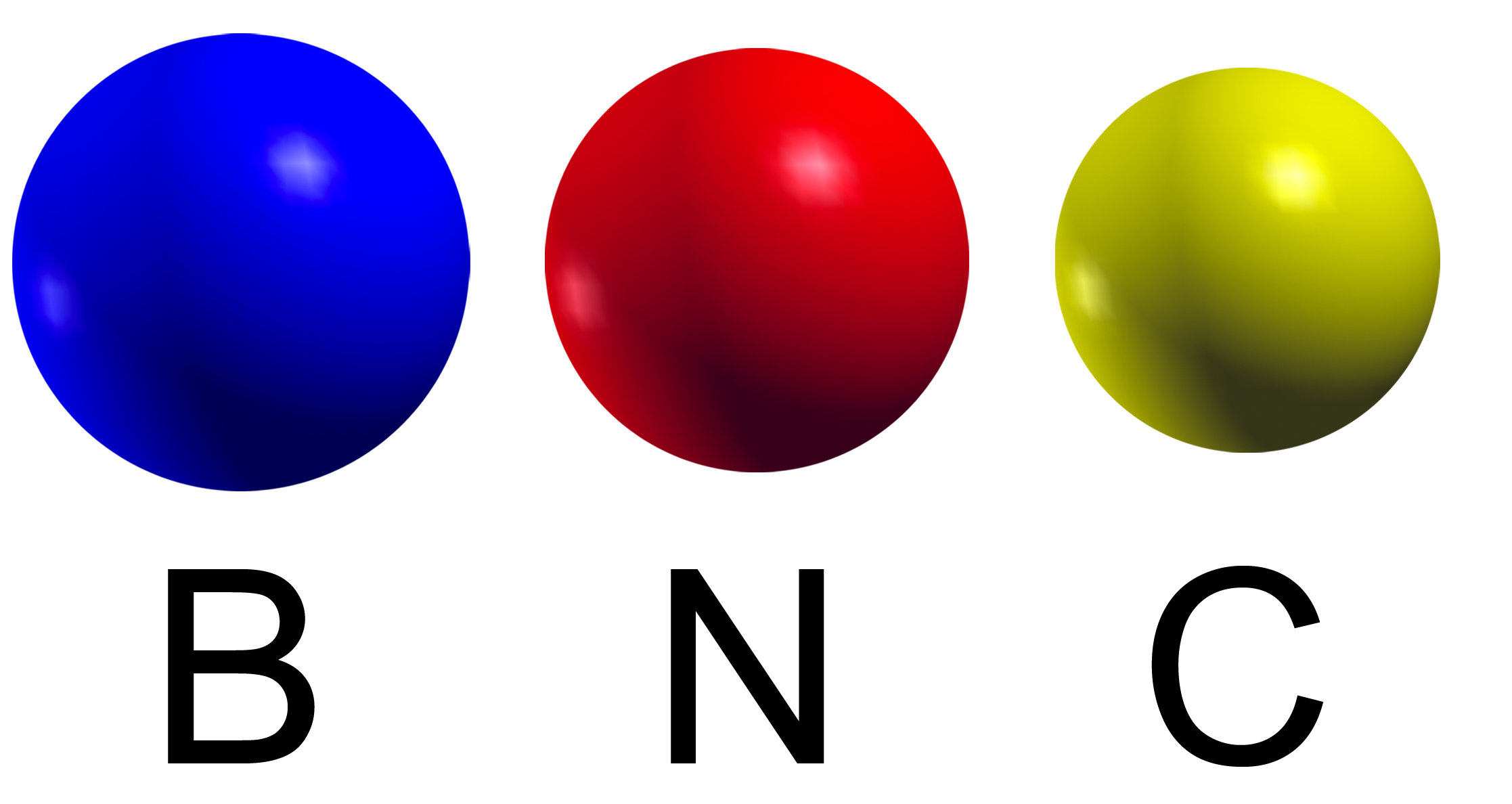}
\caption{Different ways of stacking of graphene on top of h-BN: a) AA stacking; b) AB-I stacking; b) AB-II stacking. See full description in the text.} 
\label{fig:stack}
\end{figure}

Let us, at the moment, ignore the small (about 1.7 $\%$) mismatch between the graphene and h-BN lattices. The three most symmetric stacking variations are AA, AB-I, and AB-II: for AA stacking, Fig. \ref{fig:stack}a, each carbon atom of one sublattice of graphene is on top of a boron atom of h-BN, while each carbon atom of the other sublattice of graphene is on top of the nitrogen atom of h-BN; 
for AB-I staking, Fig. \ref{fig:stack}b, each carbon atom of one sublattice of graphene is on top of a boron atom of h-BN, while each carbon atom of the other sublattice of graphene is on top of the center of a hexagon of h-BN; for AB-I staking, Fig. \ref{fig:stack}c, each carbon atom of one sublattice of graphene is on top of nitrogen atom of h-BN, while each carbon atom of the other sublattice of graphene is on top of the center of a hexagon of h-BN. First principle calculations \cite{PhysRevB.76.073103,slawinska2010energy} show that AB-II stacking configuration has a much larger energy and thus can be ignored. For the rest of the paper we will use AB to refer to AB-I stacking.

Although, the precise stacking control of graphene on h-BN is problematic, Wei Yang {\it et al} \cite{yang2013epitaxial} showed  that epitaxial growth of graphene on h-BN may allow for pure selectional stacking in this heterostructure. 

Different stacking types will affect the electrical properties of the substrate, such as the dipole moment, differently. AB stacking provides larger asymmetry between boron and nitrogen environments of h-BN lattice than AA stacking. This difference is taken into account in our calculations.

\begin{figure}
\centering
\includegraphics[width=0.8\columnwidth]{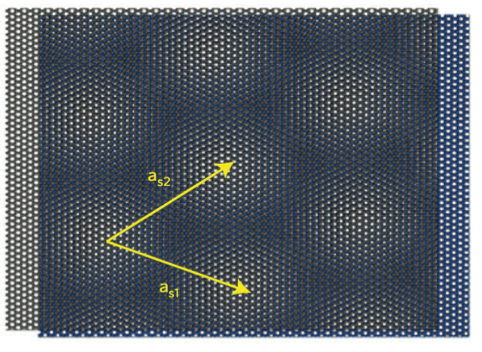}
\caption{Creation of moir\'e pattern in the presence of lattice mismatching between graphene and h-BN sublattice \cite{yang2013epitaxial}} 
\label{fig:flower}
\end{figure}

Consider now a small lattice mismatching (1.7 $\%$) between graphene and h-BN lattices. This mismatch will create a large moir\'e pattern as seen in Fig. \ref{fig:flower}. Given small lattice mismatching and small twisting angle (in the case where we do not have pure AB or AA-stacking), we can divide the supercell into two main stacking domains: AA-stacking at the corners and AB-stacking at 1/3 and 2/3 of the long diagonal \cite{trambly2010localization}. These domains fill most of the area inside the supercells. For each domain, we will have a preference site for hydrogen adsorption on just one sublattice related to the stacking of the domain. 

\section{Hydrogen Adsorption}
\label{sec:stability}
In this section, we first show that in the graphene h-BN heterostucture for both stacking AA and AB, the hydrogen is adsorbed predominantly on one sublattice of the graphene lattice. The half hydrogenated graphene will then normally become perfect graphone. We show that this graphone is stable with respect to hydrogen migration and desorption. We compare the energy differences, migration barriers, migration energies and binding energies of AA and AB stacking with that of pristine graphone.        

\begin{figure}[h]
\centering
\includegraphics[width=1\columnwidth]{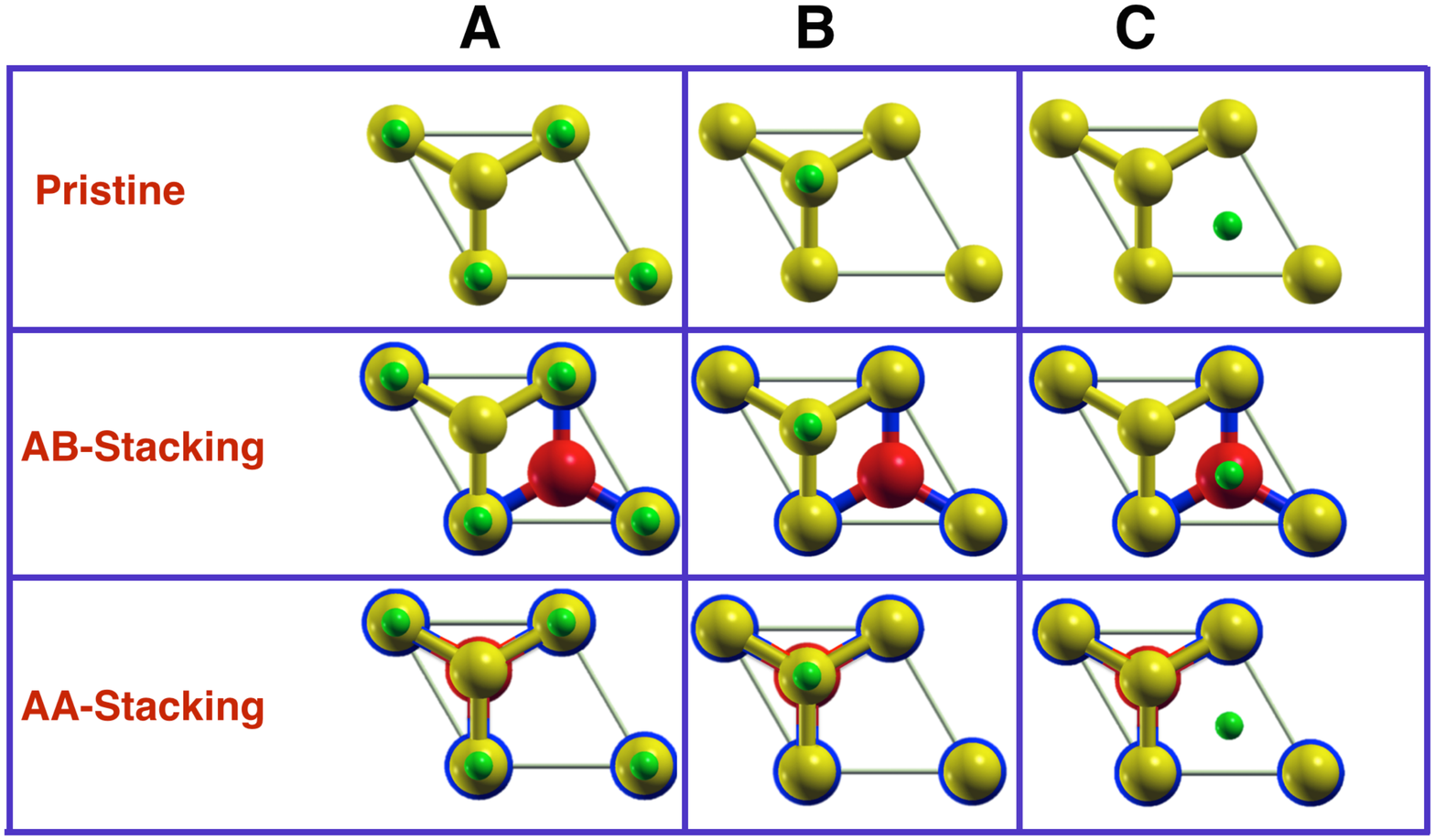}
\includegraphics[width=0.2\columnwidth]{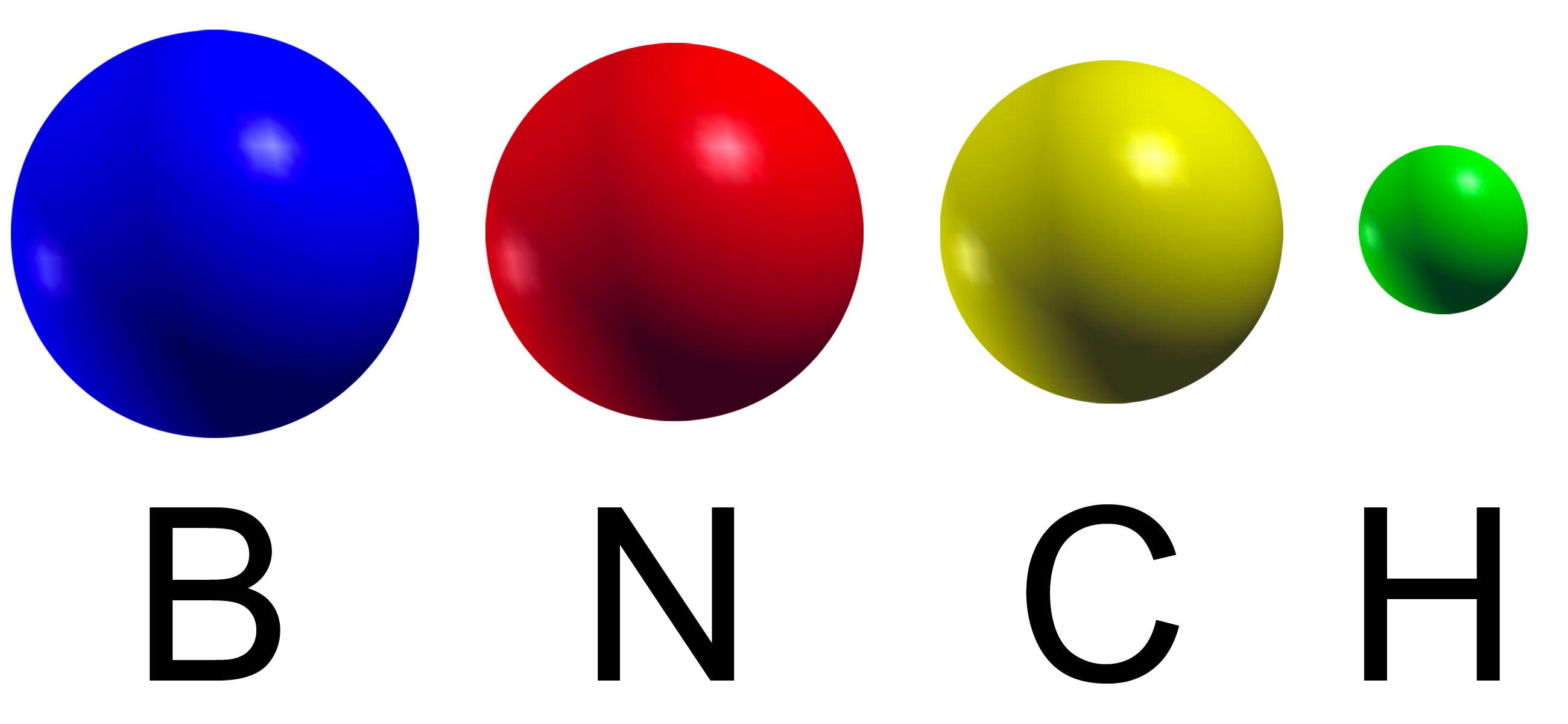}
\caption{The first row shows different hydrogenations of graphene, the second row shows  different hydrogenations of AB stacking, and the third shows that for AA stacking. Hydrogenation differs by the placement of the hydrogen atom within the unit cell. The first column is graphone(A), the second is graphone(B) and the third is graphone(C). We refer to them in the text as AB-C -- this is AB stacked heterostructure which forms graphone(C).}
\label{fig:2}
\end{figure}

There are three high symmetry adsorption sites per unit cell for pristine graphone and for any stacking types of the heterostucture. The hydrogen atom can be adsorbed on the sublattice A, sublattice B, or in the center of the hexagon C of graphene, see Fig. \ref{fig:2}. 

We use LDA, GGA, and VDW methods to find the energy difference, $\Delta E_{\mbox{\textsc{ba}}}=E_{\mbox{\textsc{a}}}-E_{\mbox{\textsc{b}}}$ and $\Delta E_{\mbox{\textsc{bc}}}=E_{\mbox{\textsc{c}}}-E_{\mbox{\textsc{b}}}$, between different graphones, A, B, and C, for both stacking types and for the pristine case. The results are shown  in the Table \ref{table:1}.   

\begin{table}[ht]
\begin{tabular}{| m{1.1cm} | m{2.5cm} | m{2.5cm} |}
\hline
Config.&\multicolumn{1}{c|}{$\Delta E_{\mbox{\textsc{ba}}}$}&\multicolumn{1}{c|}{$\Delta E_{\mbox{\textsc{bc}}}$} \\
\hline
&LDA GGA VDW&LDA GGA VDW\\
Pristine& \multicolumn{1}{c|}{0}& \multicolumn{1}{c|}{0.02}\\
AB&0.02 \space\space 0.01 \space\space 0.03 &0.03 \space\space 0.02 \space\space 0.05\\
AA&0.01 \space\space0.02 \space\space 0.02 & 0.03 \space\space 0.07 \space\space 0.08\\
\hline
\end{tabular} 
\caption{Energy differences (in eV) per unit cell, $\Delta E_{\mbox{\textsc{ba}}}=E_{\mbox{\textsc{a}}}-E_{\mbox{\textsc{b}}}$ and $\Delta E_{\mbox{\textsc{bc}}}=E_{\mbox{\textsc{c}}}-E_{\mbox{\textsc{b}}}$ for different types of graphone, A , B, C, pristine graphone and for different kinds of stacking AB and AA. For three different types of configurations (AA, AB, and pristine graphone) using LDA, GGA, and VDW approximations.} 
\label{table:1}
\end{table} 

For pristine graphone, where there are no inter-planar VDW interactions, our results reproduce the results of the previous work \cite{boukhvalov2010stable} which used GGA calculation. It is also no surprise that the graphone(A) and graphone(B) have the same energy, as in the pristine case there is no difference between them. The energy of a pristine graphone(C) is larger then the energy of pristine graphone(A) or (B).

For the graphene h-BN heterostructure the results are different. In both AA and AB stacking there is no symmetry between different sublattices. From Table \ref{table:1}, we see that the graphone(B) where hydrogen adsorbs to the sublattice B (the sublattice where is not on top of the boron atoms for both AA and AB stacking types), has the smallest energy for both ways of stacking. All three methods LDA, GGA, and VDW also show that for both kinds of stacking, graphone(C) has a larger energy than graphone(A).

The conclusion is that the site shown in the column B of Fig. \ref{fig:2} is the preference site for hydrogen adsorption. If the hydrogen atoms adsorb to all sites B on the hexagonal lattice, the result is pure graphone(B). 

We next check the stability of this graphone. First, we check graphone(B) for both ways of stacking against removal of one hydrogen atom or two neighboring hydrogen atoms. The results of LDA, GGA, and VDW calculations for both kinds of stacking as well as for pristine graphone is shown in Table \ref{table:3}.   

\begin{table}[ht]
\begin{tabular}{ |m{1.1cm} | m{2.5cm} | m{2.5cm} | }
\hline
Config.&H-Binding energy&Removal of two H\\
\hline
&LDA GGA VDW&LDA GGA VDW\\
Pristine& \multicolumn{1}{c|}{1.15} & \multicolumn{1}{c|}{-5.31}\\
AB&1.10 \space\space 1.01 \space\space 1.12&0.32 \space\space 0.36 \space\space 0.38\\
AA&1.09 \space\space 1.01 \space\space1.10& 0.31 \space\space 0.33 \space\space 0.34\\
\hline
\end{tabular} 
\caption{ Binding energy for one and two neighboring hydrogen atoms  for graphone(B) for pristine, AA, and AB stacking variations. The energies are given in eV per unit-cell.} 
\label{table:3}
\end{table} 

The conclusion is that graphone(B) is stable against desorption of hydrogen for both AA and AB stacking types of the h-BN/graphene heterostructure. We note that the pristine graphone is not stable against desorption of two neighboring hydrogen atoms.  

Finally, we check the stability of graphone(B) against migration of the hydrogen atom to a nearest site on another sublattice. In order to do that, we calculate both the migration energy and migration barrier. We denote these graphone types as graphone(A), graphone(B) and graphone(C).

For finding the migration energy, one hydrogen atom is moved per supercell ($5\times5$ unit-cell) to the nearby sublattice and then the energy difference per unit-cell for these two configurations is found. The migration barrier has been derived by the minimum-energy path calculation under the NEB method.\cite{PhysRevB.80.085428} NEB results for four different supercells, $2\times2$
, $3\times3$, $4\times4$ and $5\times5$ indicate a fairly small (less than 5 $\%$) correlation effect of two distorted adjacent supercells. For pristine graphone, only the GGA calculation was performed and results are in agreement with the previous work. \cite{boukhvalov2010stable}

\begin{table}[ht]
\begin{tabular}{ |m{1.2cm} | m{2.5cm} | m{2.5cm} | }
\hline
Config.&Migration barrier&Migration energy\\
\hline
        &\multicolumn{1}{c|}{  LDA (NEB)}&LDA GGA VDW\\
Pristine&\multicolumn{1}{c|}{0.06}&\multicolumn{1}{c|}{-1.44} \\
AB      &\multicolumn{1}{c|}{0.18}&-0.87 \space-1.04 \space -0.64\\
AA      &\multicolumn{1}{c|}{0.12}&-0.93 \space -1.12 \space-0.82 \\
\hline
\end{tabular} 
\caption{Migration barrier and migration energy for three different types of configurations.} 
\label{table:4}
\end{table} 

It was previously shown \cite{boukhvalov2010stable} that the small migration barrier makes the pristine graphone unstable against hydrogen migration. Indeed we see in Table \ref{table:4} that this barrier is just $0.06$ eV. The table also shows a substantial increase in the migration barrier for both AA and AB stacking. This increase is due to the screening effect of h-BN.

Such barriers will decrease the mobility of hydrogen atoms on top of the graphene layer in the presence of h-BN. If we start with half hydrogenated graphene, from the Arrhenius equation, $e^{(-\frac{\Delta E}{k_{B}T})}$, at room temperature, the transition probability will be less than 0.001 and from $\tau \sim \frac{\hbar}{k_{B}T}e^{(\frac{\Delta E}{k_{B}T})}$ the naive estimate for transition time will be $~10^{-10}$ sec. The transition time increases rapidly by decreasing temperature whereas at $T=60 K$, the transition time reaches to the order of 1 sec.

\begin{figure}[!h]
\centering

\includegraphics[width=0.35\columnwidth]{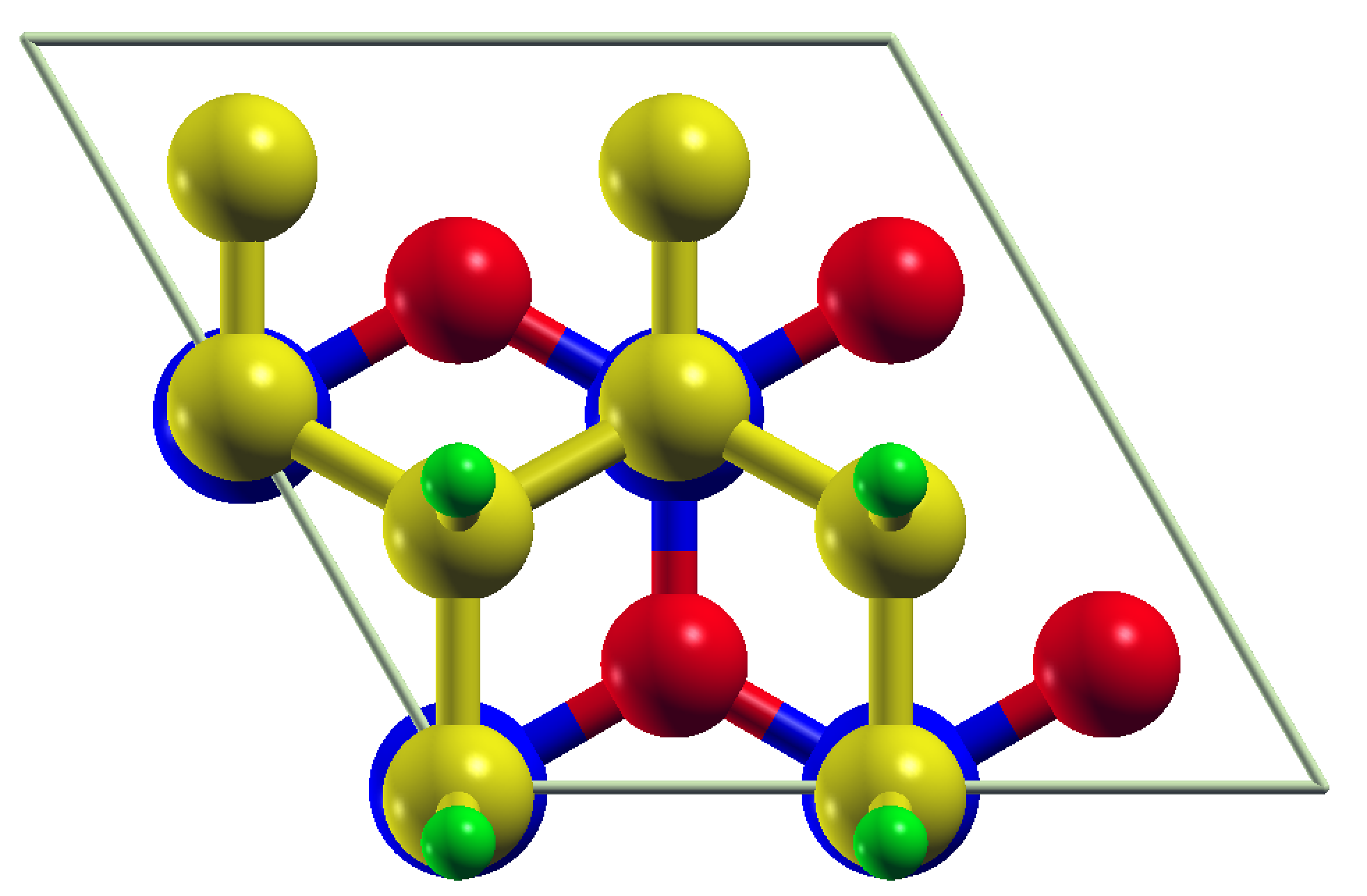}
\includegraphics[width=0.2\columnwidth]{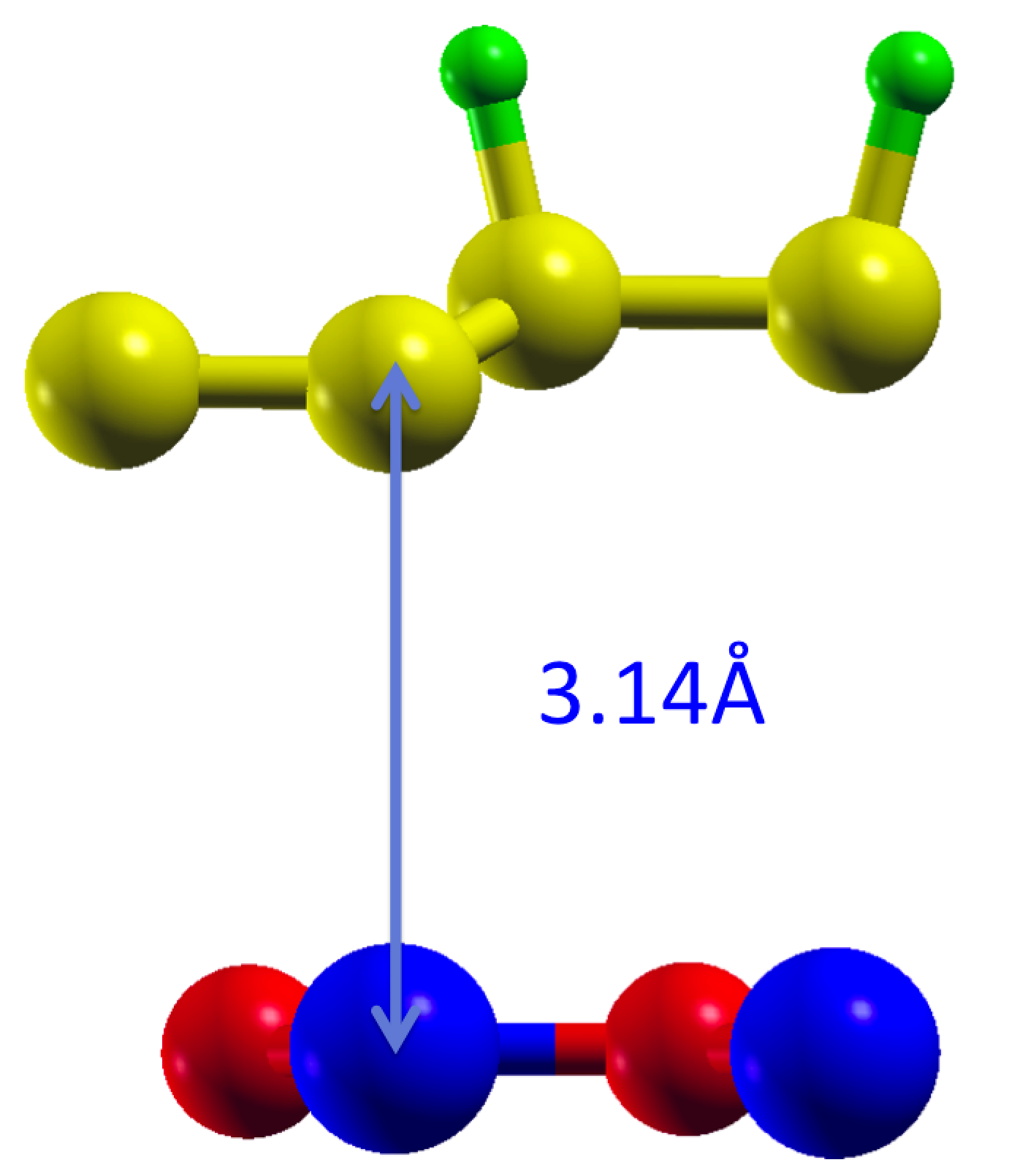}
\caption{Optimized structure for boat graphone/h-BN with 2$\times$2 supercell.}
\label{fig:boat}
\end{figure}

The migration energy for both types of stacking is still negative, although substantially less than for  pristine graphone. Although this still makes the boat graphone (see Fig. \ref{fig:boat}) the most stable configuration thermodynamically, if one exposes the graphene/hBN to the hydrogen plasma, graphone(B) will be the most stable case kinematically. This can be understood from the fact that carbon atoms belonging to sublattice B come first in contact with hydrogen atoms and they then encounter a large migration barrier due to the presence of the substrate. Moreover, it has been shown recently \cite{kharche2011quasiparticle, pumera2013graphane} that half-hydrogenated graphene only in one sublattice can be fabricated within the selective desorption of hydrogen atoms from one side of the chair graphone \cite{PhysRevB.75.153401}. Such produced graphone then will be stabilized by the high migration barrier for the h-BN/graphene heterostructure.

\section{Magnetic properties of Graphone/h-BN}\label{sec:Magnetic}
One of the main reasons why pristine graphone(A) or graphone(B) are interesting and promising is the magnetic properties of these materials.  It has been shown by spin-polarized first principle calculation in LSDA scheme that graphone (A or B) has about 1 $\mu_{B}$ per unit-cell magnetization \cite{zhou2009ferromagnetism}. Unfortunately, pristine graphone is not stable enough to be of any practical use\cite{}.  

In the previous section, we showed that h-BN substrate considerably increases the graphone(B) stability, making it feasible for further studies. It is, however, important to show that such stabilized graphone still has the magnetic properties expected from the pristine graphone. In this section, we present the results of our LSDA with GGA calculations for graphone(B) on h-BN substrate.

\begin{figure}[h]
\centering
\includegraphics[height=2.7cm]{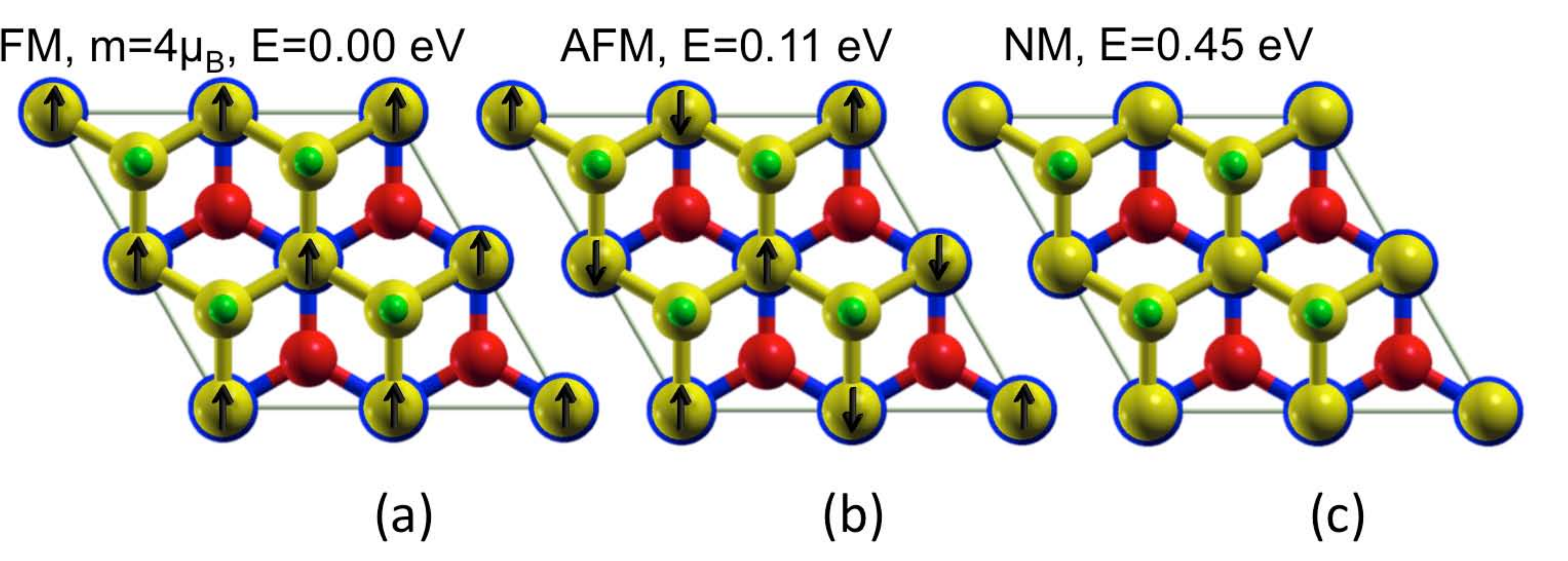}
\caption{Different magnetization states of graphone(B)/h-BN with their relative energy: a) ferromagnetic with 1 $\mu_{B}$ per unit-cell magnetization which is the most stable configuration, b) AFM state with zero magnetization, and c) normal state.} 
\label{fig:mag}
\end{figure}

Our calculations were done with fully relativistic PBE-GGA pseudo potential with non-linear core corrections and cold smearing with 0.001 eV degauss. Our supercell consists of $2\times2$ unit-cells with 20 $\AA$ vacuum layer and $16\times16\times1$ Monkhorst-Pack k-points mesh. The result shows about 1 $\mu_{B}$ magnetization and a 0.42 eV (per unit-cell) energy difference between the ferromagnetic state (FM) and the normal state, see Fig. \ref{fig:mag}. 

\begin{figure}[h]
\centering
\includegraphics[width=0.8\columnwidth]{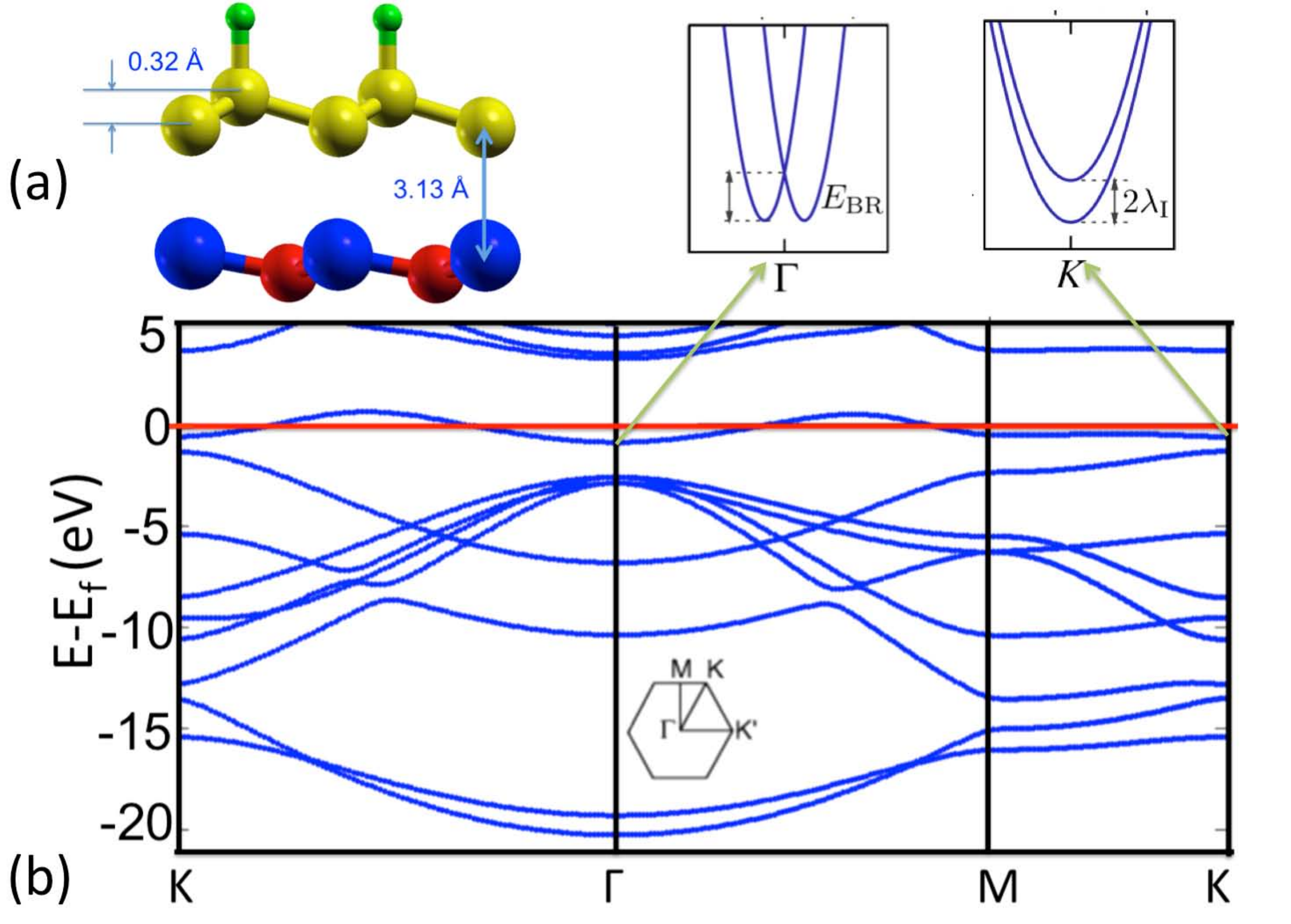}
\caption{a) Optimized structure for graphone(B)/h-BN b) Band structure of graphone/h-BN (fully relativistic GGA) with considering SOC.}
\label{fig:bs}
\end{figure}

The band structure of the optimized graphone(B)/h-BN is shown in Fig. \ref{fig:bs}. Group symmetry near K-point and $\Gamma$-point is $C_3$ and $C_{3\nu}$ respectively, whereas the  $\pi^{*}$ band is located in the vicinity of the fermi energy.\cite{PhysRevLett.110.246602} Up to the first order in momentum near the K-point and  $\Gamma$-point, these bands are described by the following SOC Hamiltonians:

\begin{equation}
H_{\mbox{\textsc{soc}}}^{\tau K}=\lambda^{\mbox{\textsc{br}}} (k_{x} s_{y}-k_{y} s_{x}) + \tau \lambda^{\mbox{\textsc{i}}} s_{z}
\label{soc}
\end{equation}

\begin{equation}
H_{\mbox{\textsc{soc}}}^{\Gamma}=\lambda^{\mbox{\textsc{br}}} (k_{x} s_{y}-k_{y} s_{x}) 
\label{soc}
\end{equation}

The two terms in Eq. \eqref{soc}, correspond to induced Bychkov-Rashba-like SOC with $\lambda^{\mbox{\textsc{br}}}=0.1$ meV and intrinsic SOC with $\lambda^{\mbox{\textsc{i}}}=1.2$ meV, respectively. $\tau$ is $+1$  ($-1$) for the K (K') point.  
$\lambda^{\mbox{\textsc{i}}}$ is half of the band splitting of $\pi^{*}$ band in the vicinity of fermi energy at the K point ($2.4$ meV) and $\lambda^{\mbox{\textsc{br}}}$ was obtained at the $\Gamma$ point. 

The increase of $\lambda^{\mbox{\textsc{i}}}$ in comparison to $\lambda^{\mbox{\textsc{i}}}=12 \mu eV$ for graphene \cite{PhysRevB.82.245412} is due to the bending of the carbon bonds, see Fig. \ref{fig:bs}. This bending changes the $sp^{2}$, plane graphene, hybridization to $sp^3$, diamond like, hybridization. This value of $\lambda^{\mbox{\textsc{i}}} $ is comparable to fully hydrogenated graphene in both sides (chair graphane \cite{zhou2010enhanced}), which reported to have 8.7 meV band splitting.

\section{TMR device: graphone/h-BN/graphone}\label{sec:TMR}

 Tunneling magneto resistance occurs in magnetic tunnel junctions that consist of two ferromagnetic metallic layers with a good insulator in between. The tunneling current is then controlled by switching magnetization in one of the layers. It is important for these devices that the ferromagnetic layers have highly spin polarized electronic states near the fermi energy.    
In this section, we show that graphone/h-BN multilayer heterostructures possess nearly 100\% polarized states at the fermi energy and thus are perfect half metals. We also show that the magnetic properties of the graphone can be controlled and enhanced by changing the number of layers of h-BN. h-BN has a large band gap (6 eV) and can be a good insulator for fabrication of TMR devices.
 These properties of the multilayer graphene h-BN heterostructures can open new horizons in spintronics.

We perform the same LDA, GGA, and VDW structure calculation for two layers of graphone(B) with 1, 2, and 3 layers of AA-stacked h-BN sandwiched in between. The results are shown in Table \ref{table:2}. 

\begin{table}[!h]
\centering
\begin{tabular}{|c|cc|cc|cc|}
\hline 
Structure & LDA& & GGA & & VDW & \\
&$d_{CB}$ & $d_{CC}^{\perp}$ &$ d_{CB} $&$ d_{CC}^{\perp}$ &$ d_{CB}$ &$ d_{CC}^{\perp}$ \\ 
\hline
graphone & 3.12 & 0.38 & 3.13 & 0.30 & 3.13 & 0.32 \\ 
g/1bn/g & 2.60 & 0.28 & 2.72 & 0.29 & 3.19 & 0.30 \\ 
g/2bn/g & 1.95 & 0.38 & 2.10 & 0.38 & 3.08 & 0.30 \\ 
g/3bn/g & 2.48 & 0.28 & 2.47 & 0.30 & 3.01& 0.31\\ 
\hline
\end{tabular} 
\caption{Optimized structure (in \AA): $d_{CB}$ is the distance between carbon on top of the boron atom and $d^{\perp}_{cc}$ is the vertical distance between two carbon atoms which belong to the same layer. It shows the magnitude of the bond bending.} 
\label{table:2}
\end{table}

\begin{figure}[!h]
\hfill
\center
\centerline{\includegraphics[height=2.6cm, scale=0.12]{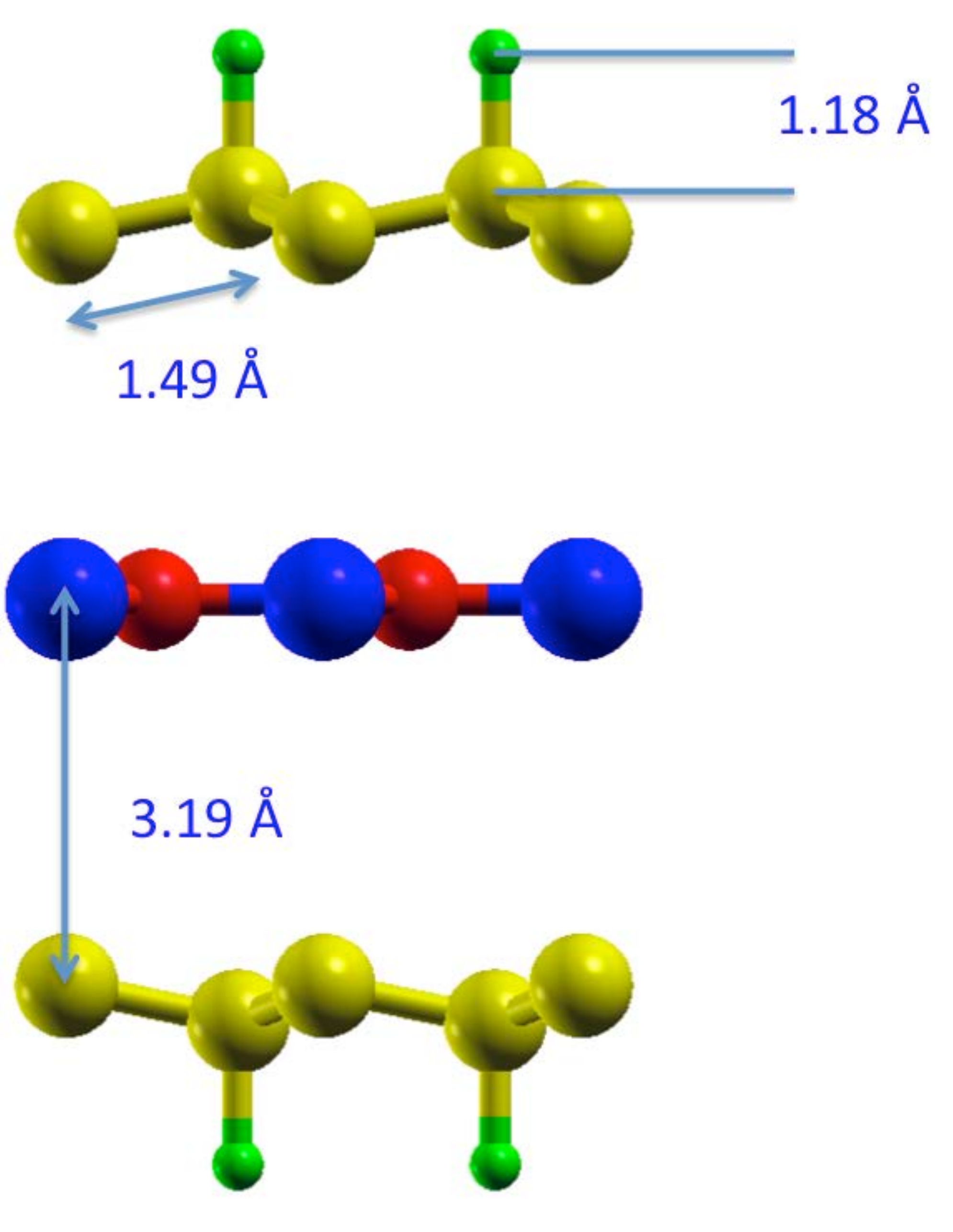}
\hfill
\includegraphics[height=3cm, width= 5.8cm]{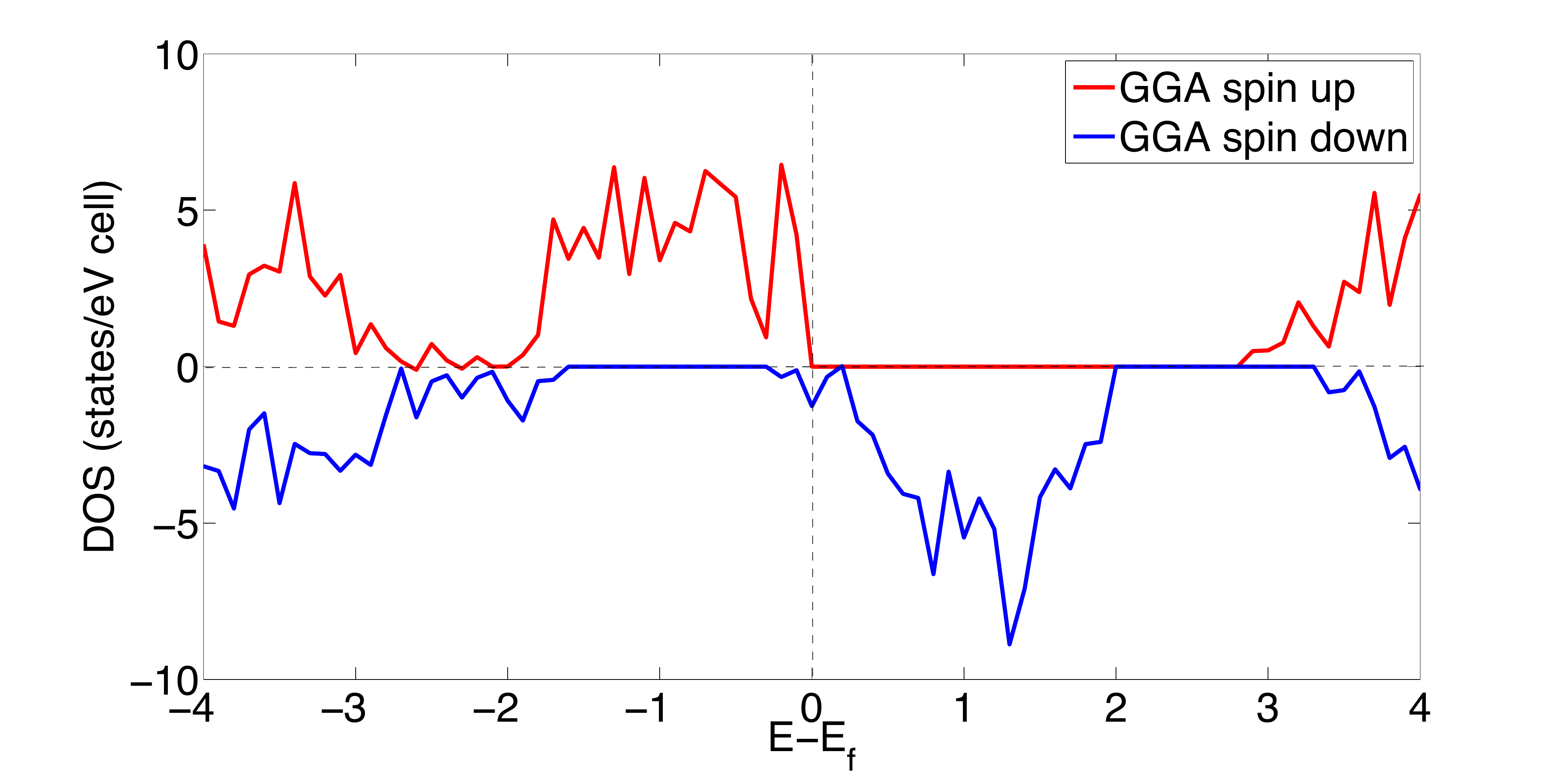}}
\vfill
\centerline{\includegraphics[height=3.3cm, scale=0.12]{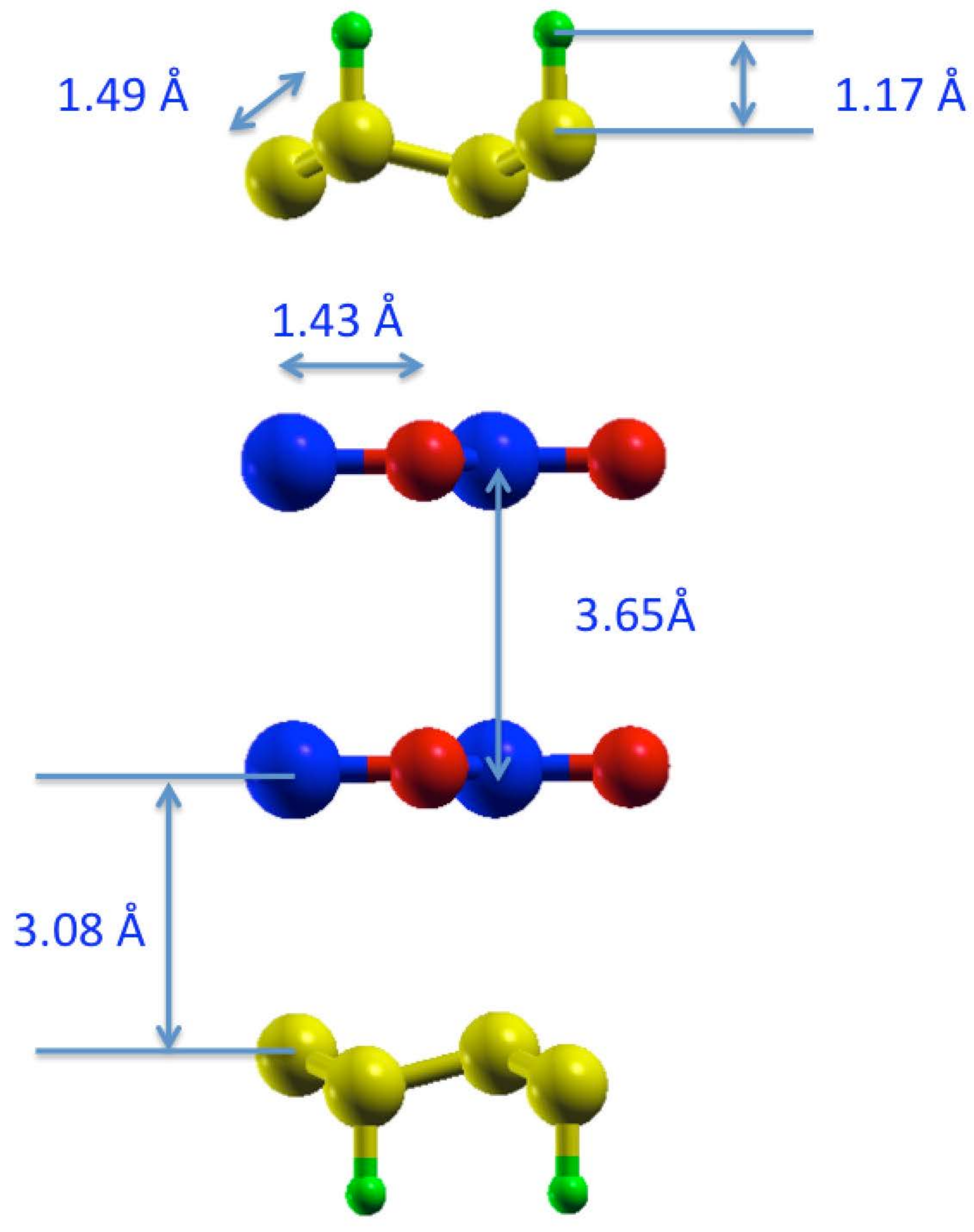}
\hfill
\includegraphics[height=3cm, width= 5.8cm]{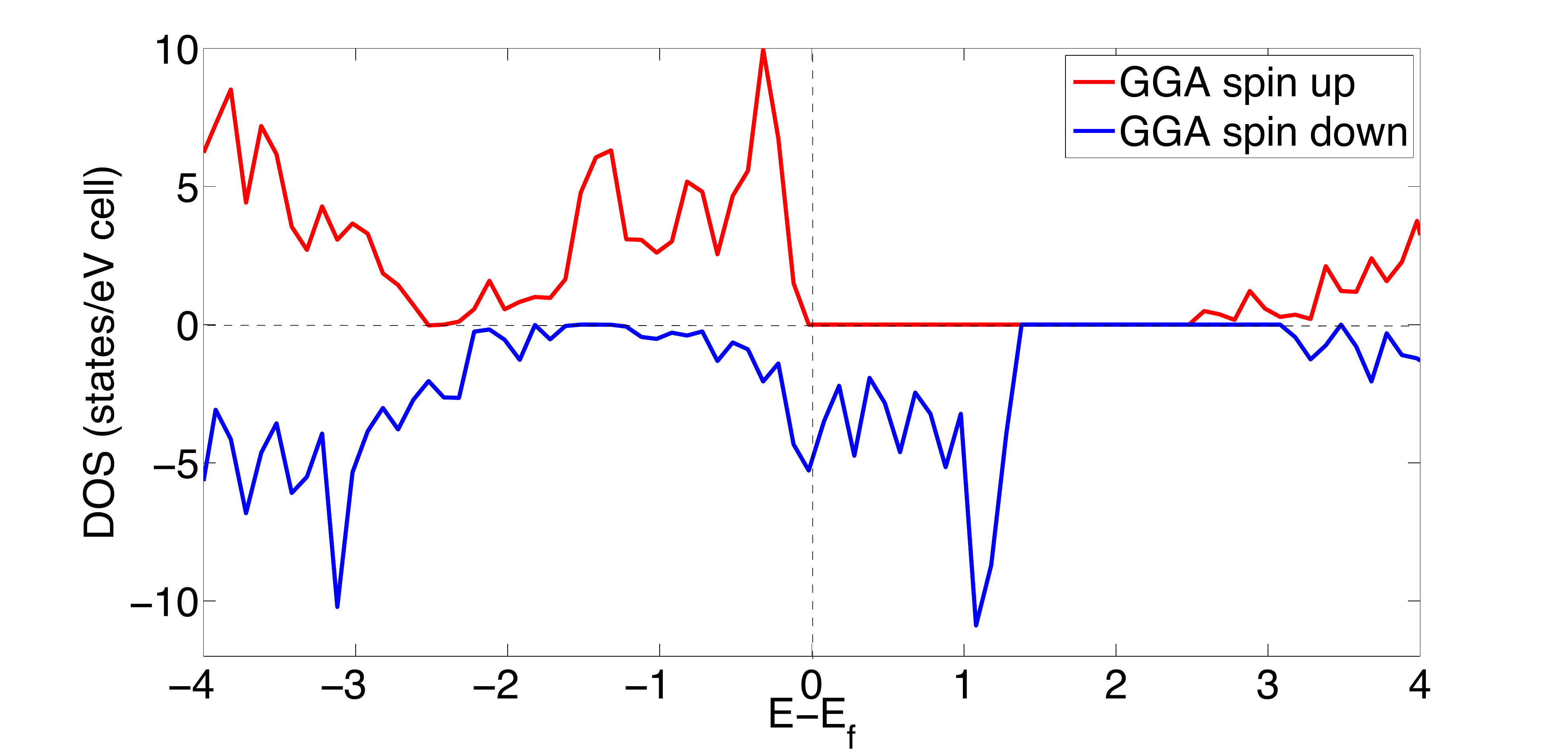}}
\vfill
\centerline{\includegraphics[height=3.8cm, scale=0.12]{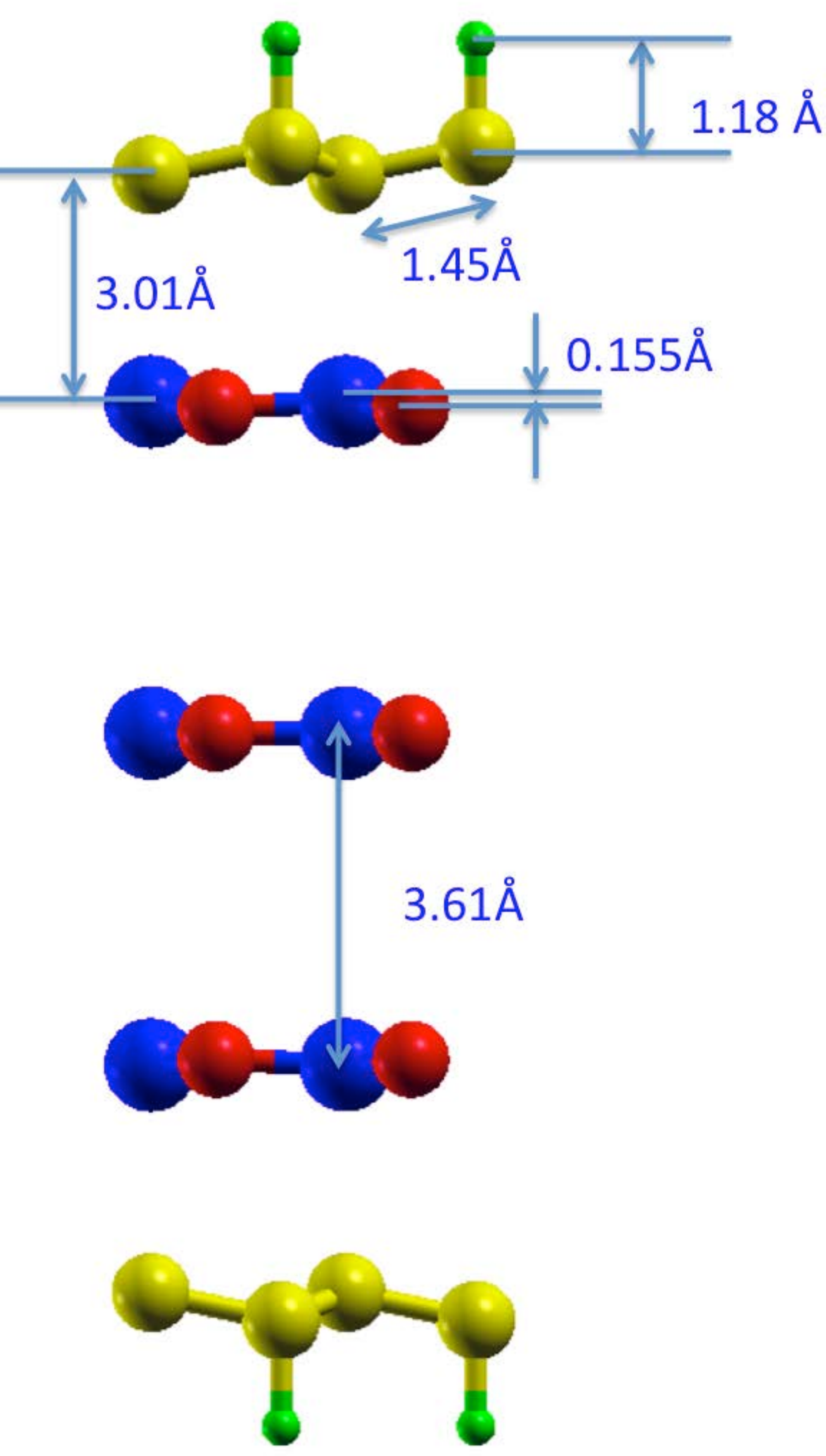}
\hfill
\hspace{0.68cm}
\includegraphics[height=3cm, width= 5.8cm]{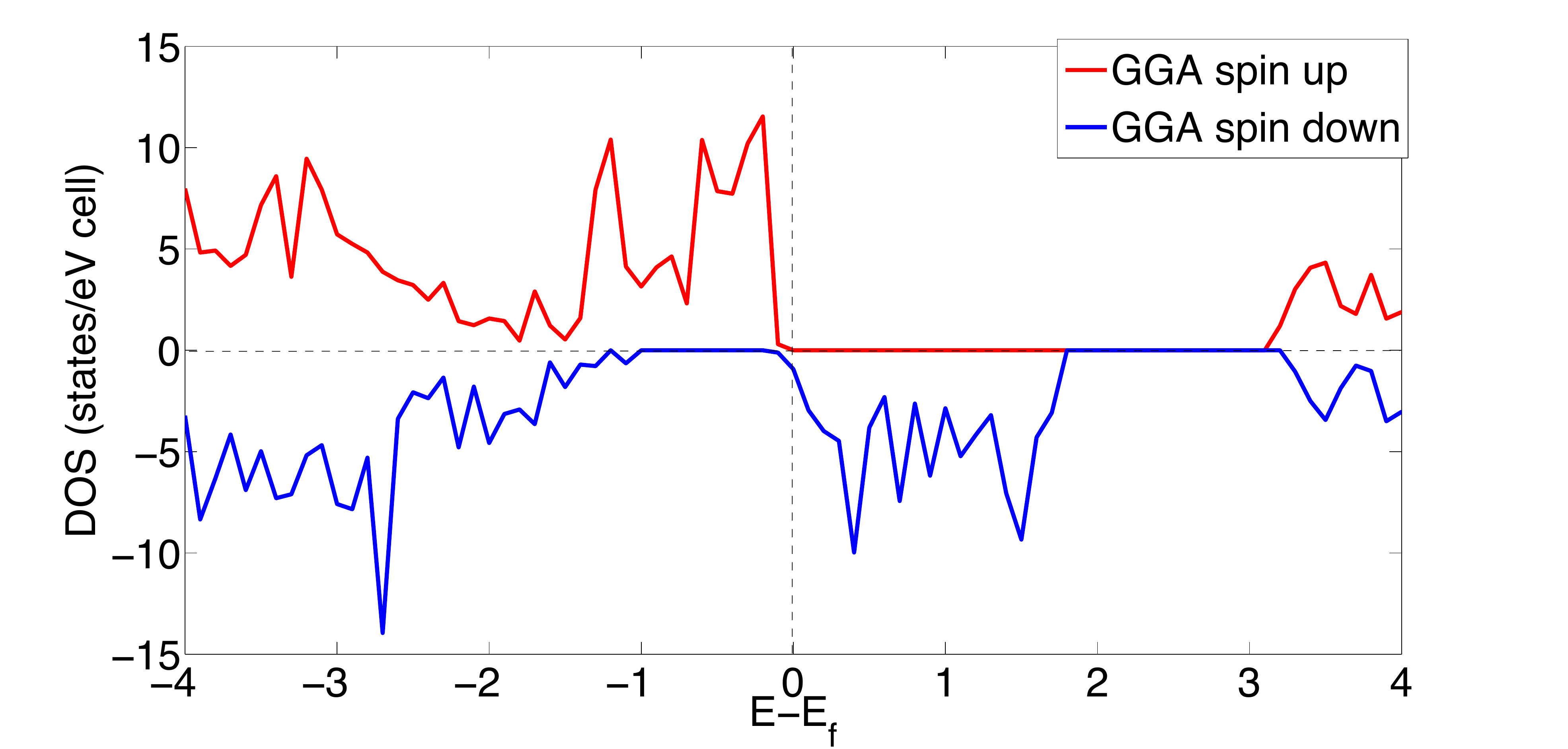}}
\caption{Optimized structure and spin-polarized DOS (all the energies are in eV) for double-layered graphone(B) and a) single , b) double and c) triple-layered AA-stacked h-BN in between.
}
\label{fig:dos}
\end{figure}

We calculated the spin polarized  density of states for the three multilayer structures using LSDA scheme with 0.01 eV degauss and cold smearing. The structural parameters for this calculations was taken from the VDW results. The results are shown in Fig. \ref{fig:dos}.

All three heterostructures show half metallic behavior near the fermi-energy. 
The spin up gap varies between 1 and 2 eV. According to Julliere model \cite{julliere1975tunneling}, such half metallic materials with 100$\%$ spin polarization are ideal for fabrication of TMR devises.

\section{Conclusion}

We proposed a feasible way to fabricate graphone in experiment by exposing the graphene/h-BN bilayer to the hydrogen plasma. The dipole moments induced by h-BN orchestrates the hydrogen pattern on the graphene layer. From first principle calculations, we have shown the presence of the preference site for hydrogen adsorption and an increment in migration barrier due to the screening effect of h-BN. The results show induced dipole moments by h-BN will trap hydrogen atoms to only one sublattice and will also kinematically stabilize the graphone layer.

The calculated band structures for the optimized graphone(B) on h-BN shows that the screening effect of h-BN not only reduces the band gap (from near 3 eV in pristine graphone to 1.93 eV in the graphone/h-BN heterostructure \cite{kharche2011quasiparticle}), but will also effectively change the fermi energy in the graphone layer.

Finally, we have shown that multilayer heterostructures (several layers of h-BN are sandwiched in between two layers of graphone) are half metal with near 100\% spin polarization. We propose to use such heterostructures in TMR devices.

We note, that it is also possible to use different elements such as fluorine as an adsorbent on graphene/h-BN heterostructures. These new materials will have different binding energies, migration barriers, and electronic properties. Doping h-BN will change the fermi energy of the graphone layer in addition to affecting hydrogen pattern on the graphene lattice. Therefore, it may provide the mechanism for control of the hydrogen pattern and electronic properties of the heterostructures.

\begin{acknowledgments}
The authors gratefully acknowledge Cristian Cernov. This work was supported by SWAN, DMR-1105512, ONR-n000141110780, and Alexander Von Humboldt
Foundation. Ar. Abanov was supported by Welch Foundation (A-1678).
\end{acknowledgments}

\bibliography{ref}
\bibliographystyle{unsrt}
\end{document}